\documentclass[12pt]{article}
\usepackage{amssymb}  

\evensidemargin \oddsidemargin
\def\1{{\bf 1}}

\def\ot{\otimes}

\def\F{\mbox{$\cal F$}}
\def\bF{\mbox{$\overline{\cal F}$}}

\def \D {{\cal D}}
\def \H {{\cal H}}
\def \W {{\cal W}}
\def \G {{\cal G}}
\def\R{\mbox{$\cal R$\,}}

\def \bV {\overline{V}_+}
\def \tW {\widetilde{W}}

\def\A{\mbox{$\cal A$}}
\def\hA{\mbox{$\widehat{\cal A}$}}

\newcommand{\tr}{\hat\triangleright}
\newcommand{\trc}{\triangleright}

\def\b#1{{\mathbb #1}}
\def\nn{\nonumber \\}

\newcommand{\be}{\begin{equation}}
\newcommand{\ee}{\end{equation}}
\newcommand{\bea}{\begin{eqnarray}}
\newcommand{\eea}{\end{eqnarray}}
\newcommand{\ba}{\begin{array}}
\newcommand{\ea}{\end{array}}
\begin{document}

\title{Can QFT on Moyal-Weyl spaces look as on commutative ones?}

\author{Gaetano Fiore  \footnote{Talk given at the 21$^{st}$ Nishinomiya-Yukawa Memorial Symposium on Theoretical Physics ``Noncommutative Geometry and Spacetime in Physics'', Nishinomiya-Kyoto, Nov. 2006.
Preprint 07-16 Dip. Matematica e Applicazioni, Universit\`a di Napoli;
DSF/12-2007. } \\ \\         \and 
        Dip. di Matematica e Applicazioni,  
        V. Claudio 21, 80125 Napoli;\\ and \\
        I.N.F.N., Sez. di Napoli, Complesso MSA, V. Cintia, 80126 Napoli 
        } 
\date{}

\maketitle

\abstract{We sketch a natural affirmative answer to the
 question based on a joint work \cite{FioWes07} with J. Wess. There
we argue that a proper enforcement of  the ``twisted Poincar\'e''
covariance makes any differences $(x\!-\!y)^\mu$ of coordinates of two 
copies of the Moyal-Weyl deformation of Minkowski space like undeformed.
Then QFT in an operator approach becomes compatible with
(minimally adapted) Wightman axioms and time-ordered perturbation theory,
and physically equivalent to ordinary QFT, as observables involve only coordinate differences. 
}

\section{Introduction: twisting Poincar\'e  group and Minkowski spacetime}

In the last decade a broad attention has been devoted to the construction
of QFT on Moyal-Weyl spaces, perhaps the simplest examples
of noncommutative spaces. These are
characterized by  coordinates $\hat x^\mu$ fulfilling
the commutation relations
\be
[\hat x^\mu,\hat x^\nu]=i\theta^{\mu\nu}, \label{cr}
\ee
where $\theta^{\mu\nu}$ is a constant real antisymmetric
matrix. For present purposes $\mu=0,1,2,3$ and indices are raised
or lowered through multiplication by the standard
Minkowski metric $\eta_{\mu\nu}$, so as to obtain a deformation of Minkowski space. 
We shall denote by $\hA$ the algebra``of functions on Moyal-Weyl space'', i.e.
the algebra generated by $\1,\hat x^\mu$ fulfilling (\ref{cr}).
For $\theta^{\mu\nu}=0$
one obtains the algebra $\A$ generated by commuting $x^{\mu}$.

Clearly (\ref{cr}) are translation invariant, but not Lorentz-covariant.
As recognized in \cite{ChaKulNisTur04,Wes04,KocTso04,Oec00}, they are
however covariant under a deformed version of the
Poincar\'e group, namely a triangular noncocommutative Hopf $*$-algebra $H$
obtained from the 
UEA $U{\cal P}$ of
the Poincar\'e Lie algebra ${\cal P}$ by {\it twisting} \cite{Dri83}\footnote{In section 4.4.1 of \cite{Oec00} this was formulated in terms 
of the dual Hopf algebra}. This means that (up to isomorphisms)
$H$  and $U{\cal P}$
(extended over the formal power series in $\theta^{\mu\nu}$)
are the same $*$-algebras, have
the same counit $\varepsilon$, but different
coproducts $\Delta,\hat\Delta$ related by
\be\ba{l}
\Delta(g)\equiv \sum_I g^I_{(1)}\ot g^I_{(2)}\:\:\longrightarrow\:\:
\hat\Delta(g)=\F\Delta(g)\F^{-1}\equiv\sum_I g^I_{(\hat 1)}\ot g^I_{(\hat
2)}\label{coproductn}
\ea\ee
for any $g\!\in\! H\equiv U{\cal P}$.
The antipodes are also changed accordingly. The socalled twist $\F$ is
not uniquely determined, but what follows does not depend on its
choice. The simplest is
\vskip-.5cm
\be\ba{l}
\qquad \qquad\quad\F\equiv \sum_I\F^{(1)}_I\ot\F^{(2)}_I:=
\mbox{exp}\left(\frac i2\theta^{\mu\nu}P_{\mu}\ot P_{\nu}\right).
                                                   \label{twist}
\ea\ee 
$P_{\mu}$ denote the generators of translations, and in (\ref{coproductn}),
(\ref{twist}), we have used Sweedler notation; $\sum_I$ may denote
an infinite sum (series), e.g. $\sum_I\F_I^{(1)}\!\ot\!\F_I^{(2)}$ comes 
out from the power expansion of the exponential. A straightforward computation  gives
$$
\hat\Delta (P_\mu)=P_\mu\!\ot\!\1\!+\!\1\!\ot\! P_\mu=
\Delta(P_\mu),\qquad\quad\hat\Delta (M_\omega)=M_\omega\!\ot\!\1\!+\!\1\!\ot\! M_\omega+
P[\omega,\theta]\!\ot\! P\neq\Delta (M_\omega),
$$
where we have set $M_\omega\!:=\!\omega^{\mu\nu}M_{\mu\nu}$ and used
a row-by-column matrix product on the right.
The left identity shows that the Hopf $P$-subalgebra remains undeformed and equivalent to the
abelian translation group  $\b{R}^4$.
Therefore, denoting by $\trc,\tr$ the actions of 
$U{\cal P},H$ (on $\A$ $\trc$ amounts to the action of
the corresponding algebra of differential operators, e.g.   $P_{\mu}$ can be
identified with $i\partial_{\mu}:= i\partial/\partial x^{\mu}$),  
they coincide on first degree polynomials in 
$x^\nu,\hat x^\nu$,
\be
P_{\mu}\trc x^{\rho}=i\delta^{\rho}_{\mu}=P_{\mu}\tr \hat x^{\rho},
\qquad\quad M_\omega\trc x^{\rho}=2i(x\omega)^\rho,
\qquad\quad M_\omega\tr \hat x^{\rho}=2i(\hat x\omega)^\rho,
   \label{Px}
\ee
and more generally on irreps (irreducible representations); 
this yields the same
classification of elementary particles as unitary irreps of ${\cal P}$. But
$\trc,\tr$ differ on products of coordinates, and more generally on tensor products of representations, as $\trc$ is extended by the rule
$g\trc\! (ab)\! =\!\big( g_{(1)}\!\trc a\big) \!\big( g_{(2)}\!\trc
b\big)$ involving $\Delta(g)$ (the rule reduces
to the usual Leibniz rule for $g=P_\mu,M_{\mu\nu}$), 
whereas $\tr$ is extended as  at the lhs of 
 \be \ba{l}
g\tr (\hat
a\hat b) =\sum_I\big( g^I_{(\hat 1)}\tr \hat a\big) \big( g^I_{(\hat 2)}\tr
\hat b\big)\quad\Leftrightarrow\quad g\trc_{\star} \!(a\!\star\! b)
=\sum_I\big( g^I_{(\hat 1)}\trc_{\star}\! a\big)\!\star\! \big( g^I_{(\hat
2)}\trc_{\star}\! b\big), \label{Leibnizn} 
\ea\ee
\vskip-2mm\noindent
involving $\hat\Delta(g)$ and a {\it deformed} Leibniz rule
for $M_\omega\tr$. Summarizing, 
the $H$-module unital $*$-algebra $\hA$ is  obtained by twisting the
$U{\cal P}$-module unital $*$-algebra $\A$.

{\bf Several spacetime variables.}
The proper noncommutative generalization of the algebra
of functions generated by $n$ sets of Minkowski coordinates
$x^{\mu}_i$, $i=1,2,...,n$, is the noncommutative
unital $*$-algebra $\hA^n$ generated by
real variables $\hat x^{\mu}_i$  fulfilling the commutation
relations at the lhs of 
\be [\hat x^{\mu}_i,\hat x^{\nu}_j]=\1 i\theta^{\mu\nu}
\quad\qquad\Leftrightarrow\quad\qquad [ x^{\mu}_i\stackrel{\star},
x^{\nu}_j]=\1 i\theta^{\mu\nu}; \label{summary} \ee
note that the commutators are not zero for $i\neq j$.
The latter are compatible with the Leibinz rule
(\ref{Leibnizn}), so as to make $\hA^n$ a
$H$-module $*$-algebra, and dictated by the braiding associated to the quasitriangular structure $\R=\F_{21}\F^{-1}$ of $H$.

As $H$ is even triangular, an essentially equivalent formulation of these
$H$-module algebras is in terms of $\star$-products derived from $\F$. 
For $n\ge 1$ denote by $\A^n$ the $n$-fold tensor product algebra
of $\A$ and $x^{\mu}\!\ot\!\1\!\ot...$, $\1\!\ot\!x^{\mu}\!\ot\!...$,... 
respectively by $x^{\mu}_1$, $x^{\mu}_2$,... Denote by $\A^n_\theta$ 
the algebra obtained by endowing the vector space underlying $\A^n$
with a new product, the $\star$-product, related to the product in $\A^n$ by
\be\ba{l}
a\star b:=\sum_I(\bF^{(1)}_I\trc a)  (\bF^{(2)}_I\trc b),  
\ea   \label{starprod}
\ee
\vskip-2mm\noindent
with $\bF\equiv\F^{-1}$. This encodes both the
usual $\star$-product within each copy of $\A$, and
the ``$\star-$tensor product'' algebra  
\cite{AscBloDimMeySchWes05,AscDimMeyWes06}.
As a result one finds the isomorphic
$\star$-commutation
relations at the rhs of (\ref{summary}) (this follows from computing
$x^{\mu}_i\!\star\! x^{\nu}_j$, which e.g. 
for the specific choice (\ref{twist}) gives 
$x^{\mu}_i x^{\nu}_j\!+\!i\theta^{\mu\nu}/2$) and that
$\hA^n,\A^n_\theta$ are isomorphic
$H$-module unital $*$-algebras, in the sense 
of the equivalence (\ref{Leibnizn}). More explicitly, on 
analytic functions $f,g$ (\ref{starprod}) reads
$f(x_i)\star g(x_j)= \exp[\frac i2\partial_{x_i}\theta\partial_{x_j}]f(x_i) 
g(x_j)$,
and must be followed by the indentification $x_i\!=\!x_j$ 
{\it after} the action of the bi-pseudodifferential operator
$\exp[\frac i2\partial_{x_i}\theta\partial_{x_j}]$ if $i\!=\!j$.
It should be
extended to functions in $L^1\cap \b{F}L^1$ in the obvious way
using their Fourier transforms $\b{F}$. In the sequel we shall formulate
the noncommutative spacetime only in terms of $\star$-products
and construct QFT on it replacing all
products of functions and/or fields with $\star$-products.

Let $a_i\!\in\!\b{R}$ with $\sum_ia_i=1$. An alternative set of 
real generators of $\A^n_\theta$ is:
\be\ba{l} \xi^{\mu}_i\!:=\! x^{\mu}_{i\!+\!1}\!-\! x^{\mu}_i,\quad
i\!=\!1,...,n\!-\!1, \qquad\quad
 X^{\mu}\!:=\!\sum_{i=1}^na_i x^{\mu}_i \quad
\ea\ee
It is immediate to check that
 $[X^{\mu}\!\stackrel{\star},\! X^{\nu}]=\1 i\theta^{\mu\nu}$, so $ X^{\mu}$
generate a copy $\A_{\theta,X}$ of $\A_{\theta}$, whereas $\forall b\!\in\!\A_{\theta}^n$
\be
\qquad \qquad\xi^{\mu}_i\star b=\xi^{\mu}_i b=b\star
\xi^{\mu}_i\qquad\Rightarrow\qquad
[\xi^{\mu}_i\stackrel{\star},b]=0, \label{startrivial} 
\ee
so $\xi^{\mu}_i$ generate a $\star$-central subalgebra
$\A_{\xi}^{n\!-\!1}$, and
$\A^n_\theta\sim\!\A_{\xi}^{n\!-\!1}\! \ot\A_{\theta,X}$.
The $\star$-multiplication operators $\xi^{\mu}_i\star$ have the same spectral decomposition on all
$\b{R}$ (including 0) as multiplication opertaors $\xi^{\mu}\cdot $
by classical coordinates, which make up a 
space-like, or a null, or a time-like $4$-vector, in the usual sense.
Moreover, 
$\A_{\xi}^{n\!-\!1},\A_{\theta,X}$ are actually $H$-module
subalgebras, with 
\be
\ba{l} g\tr a=g\trc a \qquad\qquad
\qquad\qquad a\!\in\! \A_{\xi}^{n\!-\!1},\quad g\!\in\! H  \\[8pt]
g\tr ( a\star b) \!=\!\left( g_{(1)}\trc
a\right)\!\star\! \left( g_{(2)}\tr b\right) ,
\qquad\quad b\!\in\!\A_{\theta}^n,\ea\qquad \label{UndefLeibniz} 
\ee
i.e. {\it on $\A_{\xi}^{n\!-\!1}$ the $H$-action is
undeformed}, including the related part of the Leibniz rule.
[By (10) $\star$ can be also dropped]. 
All $\xi^{\mu}_i$ are translation invariant,
$ X^{\mu}$ is not. 

\section{Revisiting Wightman axioms for QFT and their consequences}

As in Ref. \cite{Stro04} we divide the Wightman axioms \cite{StrWig63} 
into a subset (labelled by {\bf QM})  encoding
the quantum mechanical interpretation of the theory, its symmetry under
space-time translations and stability, and a
subset (labelled by {\bf R}) encoding the relativistic
properties. Since they provide 
minimal, basic requirements for the field-operator framework to quantization
we try to  apply them to the above noncommutative space keeping the QM
conditions, ``fully'' twisting Poincar\'e-covariance R1 and being ready to
weaken locality R2 if necessary.

\bigskip \noindent
{\bf QM1.} 
The states
are described by vectors of a (separable) Hilbert space $\H$.

\medskip \noindent {\bf QM2.} 
The group of space-time translations $\b{R}^4$ is represented on
$\H$ by strongly continuous unitary operators $U(a)$. The spectrum
of the generators $P_\mu$ is contained in $\overline{V}_+ = \{p_\mu: p^2
\geq 0, \,p_0 \geq 0 \}$. There is a unique  Poincar\'e invariant
state $\Psi_0$, the {\em vacuum state}.

\medskip \noindent {\bf QM3.} 
The fields (in the Heisenberg representation) $\varphi^\alpha(x)$
[$\alpha$ enumerates field species and/or $SL(2, \b{C})$-tensor
components]  are operator (on $\H$) valued tempered distributions on
Minkowski space, with $\Psi_0$ a {\em cyclic} vector for the fields,
i.e. $\star$-polynomials of the (smeared) fields applied to $\Psi_0$ give a
set $\D_0$ dense in $\H$.

\bigskip
We shall keep QM1-3. Taking v.e.v.'s we define the {\it Wightman functions}
\be 
\W^{\alpha_1,...,\alpha_n}(x_1,...,x_n):=
\left(\Psi_0,\varphi^{\alpha_1}(x_1)\star...\star\varphi^{\alpha_n}(x_n)
\Psi_0\right), 
\ee 
which are in fact distributions,  and (their combinations) the {\it Green's
functions} 
\be 
G^{\alpha_1,...,\alpha_n}(x_1,...,x_n)\!:=\!
\left(\Psi_0,T\!\left[\varphi^{\alpha_1}\!(
x_1)\!\star...\star\!\varphi^{\alpha_n}\!(x_n)\right]\!
\Psi_0\right) 
\ee  
where also {\it time-ordering } $T$ is defined
as on commutative space (even if $\theta^{0i}\neq 0$),
$$
T\!\left[\varphi^{\alpha_1}\!(x)\!\star\!\varphi^{\alpha_2}\!(y)\!\right]
\!=\!\varphi^{\alpha_1}\!(x)\!\star\varphi^{\alpha_2}\!(y)
\star\vartheta(x^0\!-\!y^0)\!
+\!\varphi^{\alpha_2}\!(y)\!\star\varphi^{\alpha_1}\!(x)
\star\vartheta(y^0\!-\!x^0)
$$ 
($\vartheta$ denotes the Heavyside function). This is well-defined
as  $\vartheta(x^0\!-\!y^0)$ is
$\star$-central. 

\medskip
QM1-3 (alone) imply exactly the same
properties as on commutative space:

\vspace{.3mm}\noindent  {\bf W1.} Wightman and Green's functions are 
translation-invariant tempered distributions
and therefore may {\it depend only on the
$\xi^{\mu}_i$}: \be\ba{rcl}
\W^{\alpha_1,...,\alpha_n}(x_1,...,x_n)&=&
W^{\alpha_1,...,\alpha_n}( \xi_1,..., \xi_{n\!-\!1}),\\
\G^{\alpha_1,...,\alpha_n}(x_1,...,x_n) &=&
 G^{\alpha_1,...,\alpha_n}( \xi_1,..., \xi_{n\!-\!1}).
\ea\ee

\noindent{\bf W2.} ({\bf Spectral condition})
The support of the Fourier transform $\tW$ of $W$ is contained in
the product of forward cones,   i.e. \be { \tW^{\{\alpha\}}(q_1,
...q_{n\!-\!1}) = 0,\qquad\mbox{if }\:\exists j:\quad q_j \notin
\bV.} \ee

\noindent{\bf W3.} $\W^{\{\alpha\}}$ fulfill
the {\bf Hermiticity and Positivity} properties following from those
of the scalar product in $\H$.

\bigskip
\noindent {\bf R1.} ({\it Untwisted} {\bf Lorentz Covariance})
$SL(2, \b{C})$ is represented  on $\H$ by strongly continuous
unitary operators $U(A)$,   and under extended Poincar\'e transformations
$U(a,A) = U(a)\,U(A)$ \be 
\qquad \qquad U(a,\! A) \,\varphi^\alpha(x)\,U(a,\!
A)^{-1}\! = S^\alpha_{\beta}(A^{-1}\!)\,\varphi^\beta\big(\Lambda(A)
x \!+\! a\big),\quad \label{transf} \ee with $S$ a finite
dimensional representation of $SL(2, \b{C})$.

\medskip
In {\it ordinary} QFT as a consequence of QM2,R1 one finds

\vspace{.5mm}\noindent {\bf W4.} ({\bf Lorentz Covariance of
Wightman functions})
 \be
 \W^{\alpha_1\!...\!\alpha_n}\! \big(\Lambda(A)x_1, ...,\Lambda(A)x_n\!\big) 
\!=\!S^{\alpha_1}_{\beta_1}(A)\!...\! S^{\alpha_n}_{\beta_n}(A)
 \W^{\beta_1\!...\!\beta_n}(x_1, ...,x_n).\quad\label{LorCov} 
\ee
In particular, Wightman (and Green) functions of scalar fields are  Lorentz invariant.

\medskip
R1 needs a ``twisted'' reformulation {\bf R1$_\star$}, which we
defer. Now, however R1$_\star$ will look like, it should imply
that $W^{\{\alpha\}}$ are $SL_\theta(2,\b{C})$
tensors (in particular invariant if all involved fields are scalar).
But, as the $W^{\{\alpha\}}$ are to be built only in terms of
$\xi^{\mu}_i$ and other $SL(2,\b{C})$ tensors (like
$\partial_{x^{\mu}_i}$, $\eta_{\mu\nu},\gamma^\mu$, etc.), which are all annihilated by
$P_\mu\trc$,   $\F$ 
will act as the identity and $W^{\{\alpha\}}$ will transform under
$SL(2,\b{C})$  as for $\theta=0$. Therefore {\bf we shall require
W4 also if $\theta\neq 0$} as a temporary substitute of R1$_\star$.

\bigskip
The simplest sensible way to formulate the $\star$-analog of locality is

\vspace{.3mm}\noindent {\bf R2}$_\star$. ({\bf Microcausality or
locality}) The fields either $\star$-commute or $\star$-anticommute at spacelike
separated points
\be { [ \,
\varphi^\alpha(x)\stackrel{\star},
\varphi^\beta(y)\,]_{\mp} = 0, \qquad\mbox{for}\,\,\,(x - y )^2 <
0.} \label{fieldcomrel} \ee
This makes sense, as space-like separation is
sharply defined, and reduces to the usual locality when $\theta=0$. 
Whether there exist reasonable weakenings of R2$_\star$ is 
 an open question also on commutative space, and the same
restrictions will apply.

Arguing as in \cite{StrWig63} one proves that  QM1-3, W4, R2$_\star$ are
independent and compatible, as they are fulfilled by free fields
(see below): the noncommutativity of
a Moyal-Weyl space is compatible with R2$_\star$!
As consequences of R2$_\star$ one again finds

\vspace{.5mm} \noindent{\bf W5.} ({\bf Locality}) if $(x_j -
x_{j+1})^2 < 0$ \be \W(x_1, ... x_j, x_{j+1}, ...x_n) =\pm \W(x_1,
...x_{j+1}, x_j, ...x_n). \ee
 \noindent{\bf W6}. ({\bf Cluster property}) For any
spacelike $a$ and for $\lambda \to \infty$ \be
 \W(x_1, ...x_j, x_{j+1} + \lambda a, ...,x_n + \lambda
a) \to \W(x_1, ...,x_j)\,\W(x_{j+1}, ...,x_n), \ee (convergence
in the distribution sense); this is true also with permuted $x_i$'s.

\medskip
Summarizing: our QFT framework is based on {\bf QM1-3, W4, R2$_\star$}, or
alternatively on the constraints {\bf W1-6} for $\W^{\{\alpha\}}$,
exactly as in QFT on Minkowski space.
We stress that this applies for all $\theta^{\mu\nu}$,
even if $\theta^{0i}\!\neq\!0$, contrary to other approaches.

\section{Free and interacting scalar field}
As the differential calculus remains undeformed, so remain
the equation of motions of free fields. Sticking for simplicity
to the case of a scalar field of mass $m$, the 
solution of the Klein-Gordon equation reads as usual
\be\ba{l}
\varphi_0(x)= \int \!d\mu(p)\,
[e^{-ip\cdot x}a^p +a_p^\dagger  e^{ip\cdot x}\,] \ea \label{fielddeco}
\ee
where $d\mu(p)=\delta(p^2\!-\!m^2)\vartheta(p^0)d^4p=
dp^0\delta(p^0\!-\!\omega_{\bf p})d^3{\bf p}/2\omega_{\bf p}$
is the invariant measure ($\omega_{\bf p}\!:=\!\sqrt{{\bf p}^2+m^2}$).
Postulating all the axioms of
the preceding section (including  {\bf R2$_\star$}),  
one can prove up to a positive factor the {\bf free field
commutation relation}
\be\ba{l}
[\varphi_0(x)\stackrel{\star},\varphi_0(y)]=
2\int \frac{d\mu(p)}{(2\pi)^3}\:\sin\left[p\!\cdot \!(x\!-\!y)\right],        \ea                \label{freecomm}
\ee
{\bf coinciding with the undeformed one}. 
Applying $\partial_{y^0}$ to (\ref{freecomm})
and setting $y^0\!=\!x^0$ [this is compatible
with (\ref{summary})] one finds {\bf the canonical commutation relation}
\be
[\varphi_0(x^0,{\bf x})\stackrel{\star},\dot\varphi_0(x^0,{\bf y})]=
i\,\delta^3({\bf x}-{\bf y}).                      \label{Cancomm}
\ee
As a consequence of (\ref{freecomm}), also the $n$-point Wightman
functions coincide with the undeformed ones, i.e. vanish if $n$ is odd  and are
sum of  products of 2-point functions (factorization)
if $n$ is even. This of course agrees with the cluster property W6.

A $\varphi_0$ fulfilling (24) can be obtained from (22) plugging
$a^p,a_p^\dagger$  satisfying
\bea
&& a^{\dagger}_p  a^{\dagger}_q= e^{i
 p\theta'\! q}\,     a^{\dagger}_q  a^{\dagger}_p, \qquad
a^p  a^q\!=\!
e^{i p\theta'\! q} \,    a^q  a^p,  \qquad
a^p a^{\dagger}_q\!=\! e^{-i  p\theta'\! q} \,  a^{\dagger}_q  a^p\!+\!2
\omega_{\bf p}\delta^3({\bf p}\!-\!{\bf q}), \nn 
&& \mbox{(with }\theta'\!=\theta\mbox{)},\qquad 
\qquad\mbox{and }[a^p,f(x)]=[a^{\dagger}_p,f(x)]=0, 
      \label{aa+cr} 
\eea 
(here $p\theta q:=p_{\mu}\theta^{\mu\nu}q_{\nu}$), as
adopted e.g. in \cite{BalPinQur05,LizVaiVit06,Abe06}.
We briefly consider the consequences of choosing $\theta'\neq\theta$
[$\theta'=0$ gives CCR among the $a^p,a_p^\dagger$,
assumed in most of the literature,  explicitly \cite{DopFreRob95}
or implicitly, in operator
\cite{ChaPreTur05,ChaMnaNisTurVer06} or in
path-integral approach to quantization].
One finds the non-local $\star$-commutation relation
$$
\varphi_0(x)\star\varphi_0(y)=e^{i
\partial_x(\theta-\theta')
\partial_y} \varphi_0(x)\star\varphi_0(y)+i\,F(x-y),
$$
and the corresponding (free field) Wightman functions violate W4,
W6, unless $\theta'=\theta$.
One can obtain (\ref{aa+cr}) also by
assuming  nontrivial transformation laws
 $P_\mu\trc a^{\dagger}_p=p_\mu a^{\dagger}_p$,
$P_\mu\trc a^p=-p_\mu a^p$ and extending the $\star$-product law
(\ref{starprod}) also to $a^p,a^{\dagger}_p$. It amounts to
choosing $\theta'\!=\!-\theta$ in (\ref{aa+cr}), see \cite{FioWes07} for details;
the relations  define examples of deformed Heisenberg algebras
covariant under a (quasi)triangular 
Hopf algebra $H$ \cite{PusWor89,Fio95}.

\noindent
{\bf Normal
ordering} is consistently defined as a map which  on any monomial 
in $a^p$, $a^{\dagger}_q$ 
reorders all  $a^p$ to the right of all $a_q^{\dagger}$ adding a factor
$e^{-i  p\theta' q}$ for each flip $a^p \leftrightarrow a_q^{\dagger}$, e.g.
$$ 
:\!a^p\! a^q\!: \:=\!a^p\! a^q, \quad
:\!a^{\dagger}_p\! a^q\!: \:=\!a^{\dagger}_p\! a^q, \quad
:\!a^{\dagger}_p\! a^{\dagger}_q\!: \:=\! a^{\dagger}_p\!
a^{\dagger}_q, \quad :\! a^p\! a^{\dagger}_q\!: \: \:=\!
a^{\dagger}_q\!  a^p e^{-i p\theta'\! q}. 
$$
(for $\theta'=0$ one finds the undeformed definition), and is extended to fields requiring $\A^n_{\theta}$-bilinearity. 
As a result, one finds that the v.e.v. of any normal-ordered 
$\star$-polynomial of fileds is zero, that normal-ordered products of
fields can be obtained from products by the same subtractions, and
{\bf the same   Wick theorem} 
as in the undeformed case. Applying
{\bf time-orderd perturbation theory} to an interacting field again one can
heuristically derive the Gell-Mann--Low formula
\be
 G( x_1,...,x_n)  =\frac{
\left(\Psi_0,T\left\{\varphi_0(x_1)\star ...\star \varphi_0(x_n)\star
\exp\left[ -i\lambda \int dy^0  \ H_I(y^0)\,\right]\right\}
\Psi_0\right)} {\left(\Psi_0,T\exp\left[ -i \int dy^0
\ H_I(y^0)\,\right]\Psi_0\right)}.
\label{formal}
\ee
Here $\varphi_0$ denotes the free ``in'' field, i.e. the incoming field in the
interaction representation, and $H_I(x^0)$ is
the interaction Hamiltonian in the interaction representation. By inspection
one finds that
the {\bf Green functions (\ref{formal}) coincide with the undeformed ones}
(at least perturbatively). They can be computed by Feynman diagrams with the
undeformed Feynman rules. See \cite{FioWes07} for some conclusions on these results, in striking contrast with the ones found in most of the literature.

\end{document}